\documentclass[runningheads,fleqn]{svmult}

\usepackage{amssymb} 

\usepackage{makeidx}   
\usepackage{graphicx}  
\usepackage{multicol}  
\usepackage{physmult}  
\makeindex             

\newcommand{\greeksym}[1]{{\usefont{U}{psy}{m}{n}#1}}
\newcommand{\umu}{\mbox{\greeksym{m}}}

\begin{document}

\title*{Self-Trapping of Light \\ 
and Nonlinear Localized Modes \\
in 2D Photonic Crystals and Waveguides}

\toctitle{Self-Trapping of Light and Nonlinear Localized Modes
\protect\newline in 2D Photonic Crystals and Waveguides}

\titlerunning{Nonlinear Localized Modes 
in Photonic Crystals}

\author{Serge F. Mingaleev $^{1,2}$ 
\and Yuri S. Kivshar $^{1}$}

\institute{$^1$ Nonlinear Physics Group, Research School of 
Physical Sciences and Engineering, \\ 
The Australian  National University, Canberra ACT 0200, 
Australia \\ 
$^2$ Bogolyubov Institute for Theoretical Physics, 
Kiev 03143, Ukraine}

\authorrunning{Serge F. Mingaleev and Yuri S. Kivshar}
\maketitle

\section{Introduction}

Photonic crystals are usually viewed as an optical analog of 
semiconductors that modify the properties of light similar
to a microscopic atomic lattice that creates a semiconductor 
band-gap for electrons \cite{book}. It is therefore believed 
that by replacing relatively slow electrons with photons as 
the carriers of information, the speed and band-width of 
advanced communication systems will be dramatically 
increased, thus revolutionizing the telecommunication 
industry. Recent fabrication of photonic  crystals with a 
band gap at optical wavelengths from 1.35  $\umu$m to 
1.95 $\umu$m makes this promise very realistic 
\cite{opt}.

To employ the high-technology potential of photonic crystals, it is 
crucially important to achieve a dynamical tunability of their 
band gap \cite{john2}. This idea can be realized by changing 
the light intensity in the so-called {\em nonlinear photonic 
crystals}, having a periodic modulation of the nonlinear 
refractive index \cite{berger}. Exploration of {\em nonlinear 
properties} of photonic band-gap (PBG) materials is an important 
direction of research that opens new applications of photonic crystals 
for all-optical signal processing and switching, allowing an 
effective way to create tunable band-gap structures operating 
entirely with light. 

One of the important physical concepts associated with 
nonlinearity is {\em the energy self-trapping and 
localization}. In the linear physics, the idea of 
localization is always associated with disorder that breaks 
translational invariance. However, during the recent years 
it was demonstrated that localization can occur in the 
absence of any disorder and solely due to nonlinearity in 
the form of {\em intrinsic localized modes} \cite{review}.  
A rigorous proof of the existence of time-periodic, 
spatially localized solutions describing such  nonlinear 
modes has been presented for a broad class of Hamiltonian 
coupled-oscillator nonlinear lattices \cite{mak},  but
approximate analytical solutions can also be found in many 
other cases, demonstrating a generality of the concept of 
{\em nonlinear localized modes}.

Nonlinear localized modes can be easily identified in 
numerical molecular-dynamics simulations in many different 
physical models (see, e.g., Ref. \cite{review} for a
review), but only very recently the {\em first experimental 
observations} of spatially localized nonlinear modes have 
been reported in mixed-valence transition metal 
complexes \cite{bishop}, quasi-one-dimensional 
antiferromagnetic chains \cite{sievers}, and arrays of 
Josephson junctions \cite{JJ}. 
Importantly, very similar types of spatially localized 
nonlinear modes have been  experimentally observed in 
{\em macroscopic} mechanical \cite{zolo} and guided-wave 
optical \cite{silb} systems.

Recent experimental observations of nonlinear localized modes, as
well as numerous theoretical results, indicate that 
nonlinearity-induced localization and spatially localized modes can be
expected in physical systems of very different nature. From the
viewpoint of possible practical applications,  self-localized states in
optics  seem to be the most promising ones; they can lead to different 
types of nonlinear all-optical switching devices where light 
manipulates and controls light itself by varying the input 
intensity. As a result, the study of {\em nonlinear
localized modes in photonic structures} is expected to bring 
a variety of realistic applications of intrinsic localized 
modes.

One of the promising fields where the concept of nonlinear 
localized modes may find practical applications is the 
physics of {\em photonic
crystals} [or photonic band gap (PBG) materials] -- 
periodic dielectric structures that produce
many of the same phenomena for photons as the crystalline 
atomic potential does for electrons \cite{book}. 
Three-dimensional (3D) photonic crystals for
visible light have been successfully fabricated only within 
the past year or two, and presently many research groups 
are working on creating tunable band-gap switches and 
transistors operating entirely with light. The most
recent idea is to employ nonlinear properties of band-gap 
materials, thus creating {\em nonlinear photonic crystals} 
including those where nonlinear susceptibility is periodic 
as well \cite{berger,sukh}. 

Nonlinear photonic crystals (or photonic crystals with 
embedded nonlinear impurities) create an ideal environment 
for the generation and observation of nonlinear localized 
photonic modes. In particular, the existence of 
such modes for the frequencies in the photonic band gaps has 
been predicted \cite{john} for 2D and 3D photonic crystals 
with Kerr nonlinearity. Nonlinear localized modes can be also 
excited at nonlinear interfaces with quadratic nonlinearity 
\cite{sukh2}, or along dielectric waveguide structures 
possessing a nonlinear Kerr-type response \cite{mcgurn}.

In this Chapter, we study self-trapping of light and 
nonlinear localized modes in nonlinear photonic crystals 
and photonic crystal waveguides. For simplicity,  we 
consider the case of a 2D photonic crystal with embedded 
nonlinear rods (impurities) and demonstrate that the 
effective interaction in such a 
structure is nonlocal, so 
that the nonlinear effects can be described by a nontrivial 
generalization of the nonlinear lattice models that include 
the long-range coupling and nonlocal nonlinearity. We 
describe several different types of nonlinear guided-wave 
states in photonic crystal waveguides and analyse their 
properties \cite{mingaleev}.  
Also, we predict the existence of {\em stable} nonlinear 
localized modes (highly localized modes analogous to gap 
solitons in the continuum limit) in the reduced-symmetry 
nonlinear photonic crystals \cite{mingaleev2}.

\section{Basic Equations}

\begin{figure}[t]
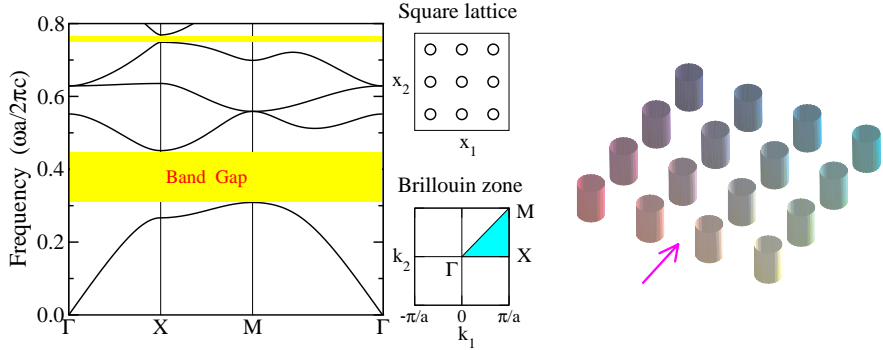

\begin{minipage}{70mm}
\includegraphics[width=70mm,angle=0,clip]{mk-fig1a.eps}
\end{minipage} 
\begin{minipage}{35mm}
\includegraphics[width=35mm,angle=270,clip]{mk-fig1b.eps}
\end{minipage} 
\vspace{3mm}
\caption{The band-gap structure of the photonic crystal 
consisting 
of a square lattice of dielectric rods with $r_0=0.18a$ and 
$\varepsilon_0=11.56$ (the band gaps are shaded). 
The top center inset shows a cross-sectional 
view of the 2D photonic crystal depicted in the right inset. 
The bottom center inset shows the corresponding Brillouin 
zone, with the irreducible zone shaded.}
\label{fig:band-r0.18}
\end{figure}

Let us consider a 2D photonic crystal created by a periodic 
lattice of parallel, infinitely long dielectric rods in
air (see Fig. \ref{fig:band-r0.18}). 
We assume that the rods are parallel to the 
$x_3$ axis, so that the system is characterized by the 
dielectric constant $\varepsilon(\vec{x})=\varepsilon(x_1,
x_2)$. As is well known \cite{book}, the photonic 
crystals of this 
type can possess a complete band gap for the $E$-polarized 
(with the electric field $\vec{E} \, || \, \vec{x}_3$) 
light propagating in the $(x_1, x_2)$-plane. 
The evolution of such a light is governed by the scalar 
wave equation
\begin{equation}
\nabla^2 E(\vec{x}, t) - \frac{1}{c^2} \, \partial_t^2
\left[ \varepsilon(\vec{x}) E \right] = 0 \; ,
\label{sys:eq-E-t}
\end{equation}
where
$\nabla^2 \equiv \partial_{x_1}^2 + \partial_{x_2}^2$ 
and $E$ is the $x_3$ component of $\vec{E}$.
Taking the electric field in the form
$E(\vec{x}, t) = \E^{- \I \omega t} \, E(\vec{x}, t \,|\, 
\omega) \; ,$ 
where $E(\vec{x}, t \,|\, \omega)$ is a slowly varying 
envelope, i.e. 
$\partial^2_t E(\vec{x}, t \,|\, \omega) \ll \omega 
\partial_t E(\vec{x}, t \,|\, \omega)$, 
Eq. (\ref{sys:eq-E-t}) reduces to 
\begin{equation}
\left[ \nabla^2  + \varepsilon(\vec{x}) 
\left( \frac{\omega}{c} \right)^2 \right]
E(\vec{x}, t \,|\, \omega) \simeq -2 \, \I \,
\varepsilon(\vec{x}) \frac{\omega}{c^2} \, 
\frac{\partial E}{\partial t} \; .
\label{sys:eq-E-omega-t}
\end{equation}
In the stationary case, i.e. when the r.h.s. of Eq. 
(\ref{sys:eq-E-omega-t}) vanishes, 
this equation describes an eigenvalue problem which can be 
solved, e.g. by the plane waves method 
\cite{Maradudin:1993:PBGL}, in the case of a perfect 
photonic crystal, for which the dielectric constant 
$\varepsilon(\vec{x}) \equiv \varepsilon_{p}(\vec{x})$ 
is a periodic function defined as
\begin{equation}
\varepsilon_{p}(\vec{x}+\vec{s}_{ij}) =
\varepsilon_{p}(\vec{x}) \; , 
\label{sys:eps-pc}
\end{equation} 
where $i$ and $j$ are arbitrary integers, and 
\begin{equation}
\vec{s}_{ij} = i \, \vec{a}_1 + j \, 
\vec{a}_2
\label{sys:s-ij}
\end{equation}
is a linear combination of the lattice vectors
$\vec{a}_1$ and $\vec{a}_2$.
 
For definiteness, we consider the 2D photonic crystal
earlier analyses (in the linear limit) 
in Refs. \cite{Mekis:1996:PRL,Mekis:1998:PRB}. That is, 
we assume that cylindrical rods with radius 
$r_0=0.18a$ and dielectric constant $\varepsilon_0=11.56$ 
form a square lattice with the distance $a$ between two 
neighboring rods, so that 
$\vec{a}_1=a \vec{x}_1$ and $\vec{a}_2=a \vec{x}_2$.
The frequency band structure for this type of 2D photonic 
crystal is shown in Fig. \ref{fig:band-r0.18} where, using
the notations of the solid-state physics, the wave dispersion
is mapped onto the Brillouin zone of the so-called 
{\em reciprocal lattice} that faces are known as $\Gamma$,
$M$, and $X$. As follows from Fig. \ref{fig:band-r0.18}, 
there exists a large (38\%) band gap 
that extends from the lower cut-off frequency, 
$\omega=0.302 \times 2\pi c/a$, to the upper band-gap 
frequency, $\omega=0.443 \times 2\pi c/a$. 
If the frequency of a low-intensity light falls into the 
band gap, the light cannot propagate through the photonic 
crystal and is reflected.

\section{Defect Modes: The Green Function 
Approach}

One of the most intriguing properties of photonic band 
gap crystals is the emergence of exponentially localized modes
that may appear  within the photonic band gaps when a defect 
is embedded into an otherwise perfect photonic crystal. 
The simplest way to create a defect in a 2D photonic 
crystal is to introduce an additional defect rod with 
the radius $r_d$ and the dielectric constant 
$\varepsilon_{d}(\vec{x})$. 
In this case, the dielectric constant 
$\varepsilon(\vec{x})$ can be presented as a sum of periodic 
and defect-induced terms, i.e. 
\begin{displaymath}
\varepsilon(\vec{x})=\varepsilon_{p}(\vec{x})+
\varepsilon_{d}(\vec{x}) \; ,
\end{displaymath}
and, therefore, Eq. (\ref{sys:eq-E-omega-t}) takes the form 
\begin{eqnarray}
\left[ \nabla^2 + \left( \frac{\omega}{c} \right)^2
\varepsilon_{p}(\vec{x}) \right] E(\vec{x}, 
t \,|\, \omega) = - \hat{\cal L} E(\vec{x}, t 
\,|\, \omega) \; ,
\label{sys:eq-E-omega-t2}
\end{eqnarray}
where the operator 
\begin{eqnarray}
\hat{\cal L} = \left( \frac{\omega}{c} \right)^2 
\varepsilon_{d}(\vec{x}) 
+ 2 \, \I \, \varepsilon(\vec{x}) 
\frac{\omega}{c^2} \frac{\partial}{\partial t} 
\end{eqnarray}
is introduced for convenience. 
Equation (\ref{sys:eq-E-omega-t2}) can also be written in 
the equivalent integral form
\begin{equation}
E(\vec{x}, t \,|\, \omega) = 
\int \D^2\vec{y} \,\,\, G(\vec{x}, 
\vec{y} \,|\, \omega) \, \hat{\cal L} \, 
E(\vec{y}, t \,|\, \omega) \; ,
\label{sys:eq-green-int}
\end{equation}
where $G(\vec{x}, \vec{y} \,|\, \omega)$ is the Green 
function defined, in a standard way, as a 
solution of the equation
\begin{equation}
\left[ \nabla^2 + \left( \frac{\omega}{c} \right)^2
\varepsilon_{p}(\vec{x}) \right]
G(\vec{x}, \vec{y} \,|\, \omega) = - 
\delta(\vec{x}-\vec{y}) \; .
\label{sys:eq-green-omega}
\end{equation}
General properties of the Green function of a perfect 2D 
photonic crystal are described in more details in Ref.
\cite{Maradudin:1993:PBGL}. Here, we notice
that the Green function is {\em symmetric}, i.e. 
\begin{displaymath}
G(\vec{x}, \vec{y} \,|\, \omega) =
G(\vec{y}, \vec{x} \,|\, \omega)
\end{displaymath}
and {\em periodic}, i.e.
\begin{displaymath}
G(\vec{x} + \vec{s}_{ij}, \vec{y}  + \vec{s}_{ij} \,|\, 
\omega) = G(\vec{x}, \vec{y} \,|\, \omega) \; ,
\end{displaymath}
where $\vec{s}_{ij}$ is defined by Eq. (\ref{sys:s-ij}).

\begin{figure}[t]
\begin{minipage}{40mm}
\centerline{{\large\bf (a)}}
\includegraphics[width=40mm,angle=0,clip]{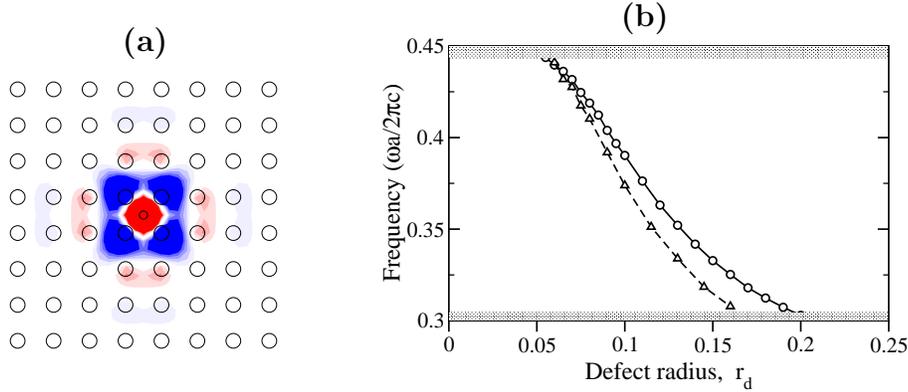}
\end{minipage} \hspace*{10mm}
\begin{minipage}{70mm}
\centerline{{\large\bf (b)}}
\includegraphics[width=70mm,angle=0,clip]{mk-fig2b.eps}
\end{minipage}
\vspace{3mm}
\caption{{\bf (a)} Electric 
field structure of a linear localized mode
supported by a single defect rod with radius 
$r_d=0.1a$ and $\varepsilon_d=11.56$ in a square-lattice
photonic crystal with $r_0=0.18a$ and
$\varepsilon_0=11.56$. 
The rod positions are indicated by circles and 
the amplitude of the electric field is indicated by color.
{\bf (b)} 
Frequency of the defect mode as a function of the 
radius $r_d$: calculated precisely from 
Eq. (\ref{sys:eq-green-int}) ({\em full line with circles}) 
and approximately from Eq. (\ref{sys:eq-defect}) 
({\em dashed line with triangles}).}
\label{fig:defect}
\end{figure}

The Green function can be calculated by means of the 
Fourier transform
\begin{equation}
G(\vec{x}, \vec{y} \,|\, \omega) = 
\int_{-\infty}^{\infty} \D t \,\,\, 
e^{\I \omega t} G(\vec{x}, \vec{y}, t)
\end{equation}
applied to the time-dependent Green function governed by 
the equation
\begin{equation}
\left[ \nabla^2  - \varepsilon_{p}(\vec{x}) \, \partial_t^2
\right] G(\vec{x},\vec{y}, t) = - \delta(t) \,
\delta(\vec{x}-\vec{y}) \; ,
\end{equation}
which has been solved by the finite-difference time-domain 
method \cite{Ward:1998:PRB}. 

Now that we have calculated the Green function, we can 
figure out the defect states solving Eq. (\ref{sys:eq-green-int})
directly. For example, Fig.~\ref{fig:defect}(a) 
shows a defect mode created by introducing a single 
defect rod with the radius $r_d=0.1a$ and dielectric constant 
$\varepsilon_d=11.56$ into the 2D photonic crystal shown 
in Fig.~\ref{fig:band-r0.18}. 
Although direct numerical solution of the integral
equation (\ref{sys:eq-green-int}) remains possible even 
in the case of a few defect rods, it becomes severely limited 
by the current computer facilities as soon as we increase 
the number of the defect rods and start investigation of
the line defects (waveguides) and their branches. 
Thus, looking for new approximate numerical techniques which 
could combine reasonable accuracy, flexibility, and power 
to forecast new effects is an issue of the key importance. 

\section{Effective discrete equations}

Studying the electric field distribution of the defect
mode in Fig. \ref{fig:defect}(a), one can suggest that 
a reasonably accurate approximation should be provided by 
the assumption that the electric field inside the defect rod
remains constant.  Indeed, let us assume that nonlinear defect rods 
embedded into a photonic  crystal are located at the points
$\vec{x}_m$, where $m$ is the index (or a combination of 
two indices in the case of a two-dimensional array of 
defect rods) introduced for explicit numbering of the 
defect rods.
In this case, the correction to the dielectric constant is
\begin{eqnarray}
\varepsilon_d(\vec{x}) = \left\{\varepsilon_{d}^{(0)} +
|E(\vec{x}, t \,|\, \omega)|^2\right\}
\sum_m \theta (\vec{x}-\vec{x}_m) \; ,
\label{sys:delta-eps}
\end{eqnarray}
where
\begin{equation}
\theta (\vec{x}) = \left\{
\begin{array}{c}
1 \; , \quad \mbox{for} \quad |\vec{x}| \leq r_d \; , \\
0 \; , \quad \mbox{for} \quad |\vec{x}| > r_d \; .
\end{array}
\right.
\label{sys:theta}
\end{equation}
The second term in Eq. (\ref{sys:delta-eps}) takes into 
account a contribution due to the Kerr nonlinearity 
(we assume that the electric field is scaled with the 
nonlinear susceptibility, $\chi^{(3)}$). 
Assuming, as we discussed above, that the electric field 
$E(\vec{x}, t \,|\, \omega)$ inside the defect rods is 
almost constant, one can derive, by substituting 
Eq. (\ref{sys:delta-eps}) into Eq. (\ref{sys:eq-green-int}) 
and averaging over the cross-section of the rods 
\cite{mcgurn}, 
an approximate {\em discrete nonlinear equation} 
\begin{eqnarray}
i \sigma \frac{\partial}{\partial t} E_{n} - E_n 
+ \sum_m J_{n-m}(\omega) (\varepsilon_{d}^{(0)} + 
|E_m|^2) E_m = 0 \; ,
\label{sys:eq-E-disc}
\end{eqnarray}
for the amplitudes of the electric field 
$E_n(t \,|\, \omega) \equiv E(\vec{x}_n, t \,|\, \omega)$
inside the defect rods. The parameter $\sigma$ and the 
coupling constants 
\begin{equation}
J_{n}(\omega) = \left( \frac{\omega}{c} \right)^2
\int_{r_d} d^2 \vec{y} \,\,\,
G(\vec{x}_0, \vec{x}_n + \vec{y} \,|\, \omega ) 
\label{sys:Jn}
\end{equation}
are determined in this case by the Green function 
$G(\vec{x}, \vec{y} \,|\, \omega)$ 
of the perfect photonic crystal. 

To check the accuracy of the approximation provided 
by Eq. (\ref{sys:eq-E-disc}),  we solved it in the 
linear limit for the case of a single defect rod. 
In this case Eq.~(\ref{sys:eq-E-disc}) 
is reduced to the equation 
\begin{equation}
J_{0}(\omega_{d})= 1/\varepsilon_{d}^{(0)} \; , 
\label{sys:eq-defect}
\end{equation}
from which one can obtain an estimation for 
the frequency $\omega_{d}$ of the localized defect mode. 
As is seen from Fig. \ref{fig:defect}(b), the mode frequency 
calculated in the framework of this approximation is in a good 
agreement with that calculated directly from  Eq.
(\ref{sys:eq-green-int}), provided the defect radius 
$r_d$ is small enough. Even for $r_d=0.15a$ an error
introduced by the approximation does not exceed 5\%.
It lends a support to the 
validity of Eq. (\ref{sys:eq-E-disc}) allowing 
us to use it hereafter for studying nonlinear localized 
modes. 

\section{Nonlinear Waveguides in 2D Photonic Crystals}

One of the most promising applications of the PBG structures
is a possibility to create a novel type of optical waveguides. 
In conventional waveguides such as 
optical fibers, light is confined by {\em total internal 
reflection} due a difference in the refractive indices of 
the waveguide core and cladding. 
One of the weaknesses of such waveguides is that 
creating of bends is difficult. Unless the radius of 
the bend is large compared to the wavelength, much of the 
light will be lost. This is a serious obstacle for creating 
``integrated optical circuits'', since the space 
required for large-radius bends is unavailable. 

\begin{figure}[t]
\centerline{
\includegraphics[width=70mm,angle=0,clip]{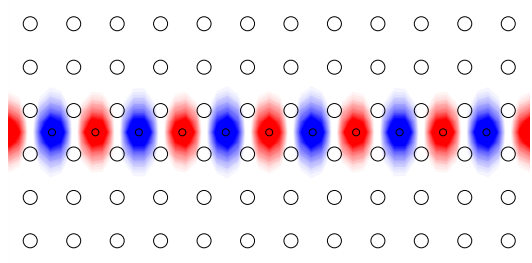}}
\caption{Electric field of the linear guiding mode 
in a waveguide created by an array of the defect 
rods. The rod positions are indicated by circles and 
the amplitude of the electric field is indicated by color.}
\label{fig:wg}
\end{figure}

The waveguides based on 
the PBG materials employ {\em a different physical 
mechanism}: the light is 
guided by a line of coupled defects which possess a 
localized defect mode with frequency inside the band gap. 
The simplest photonic-crystal waveguide can be created by 
a straight line of defect rods,  as shown in Fig. 
\ref{fig:wg}. Instead of a single localized state of 
an isolated defect, a waveguide supports propagating states 
(guided modes) with the frequencies in a narrow band 
located inside the band gap of a perfect crystal. 
Such guided modes have a periodical profile along the 
waveguide, and they decay exponentially in the transverse 
direction,  see Fig. \ref{fig:wg}.
That is, photonic crystal waveguides operate in a manner 
similar to
resonant cavities, and the light with guiding frequencies 
is forbidden from propagating in the bulk.
Because of this, when a bend is created in a 
photonic crystal waveguide, 
the light remains trapped and the only possible problem 
is that of reflection. However, as was predicted numerically 
\cite{Mekis:1996:PRL,Mekis:1998:PRB} 
and then demonstrated in microwave \cite{Lin:1998:SCI} 
and optical \cite{Tokushima:2000:APL} experiments, 
it is still possible to get very high transmission 
efficiency for nearly all frequencies inside the gap.

\begin{figure}[t]
\centerline{
\includegraphics[width=80mm,angle=0,clip]{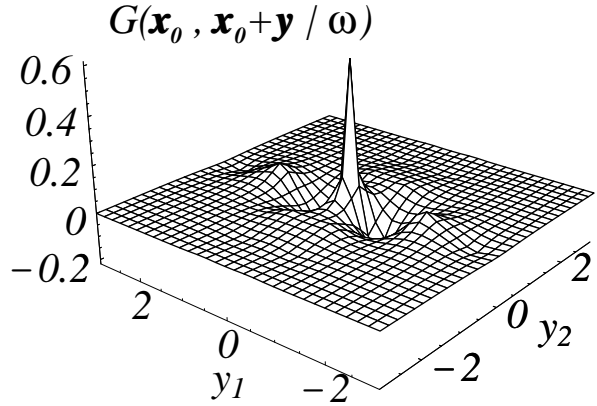}}
\caption{The Green function $G(\vec{x}_0, \vec{x}_0 + 
\vec{y} \,|\, \omega)$ for the photonic crystal shown in 
Fig.~\protect\ref{fig:band-r0.18} 
($\vec{x}_0=\vec{a}_1/2$ and
$\omega=0.33 \times 2\pi c/a$).}
\label{fig:green-r0.18}
\end{figure}

To employ the high-technology potential of photonic crystal 
waveguides, it 
is crucially important to achieve a tunability of their 
transmission properties. Nowadays, several 
approaches have been suggested for this purpose. 
For instance, it has been recently demonstrated both 
numerically \cite{Cheng:1999:PRB}
and in microwave experiments \cite{Jin:1999-sep:APL},
that transmission spectrum of straight and sharply bent 
waveguides 
in {\em quasiperiodic photonic crystals} is rather rich in 
structure and only some frequencies get near perfect
transmission. Another possibility is creation of the 
{\em channel drop system} on the bases of two parallel 
waveguides coupled by the point defects between them. 
It has been shown \cite{Fan:1998-feb:PRL} 
that high-Q frequency selective complete transfer can occur
between such waveguides by creating resonant defect states
of different symmetry and by forcing an accidental
degeneracy between them. 

However, being frequency selective, the above mentioned 
approaches do not possess {\em dynamical tunability} of 
the transmission properties. The latter idea can be realized
by changing the light intensity in the so-called 
{\em nonlinear photonic crystal waveguides} \cite{mingaleev}, 
created by inserting an additional row of rods made 
from a Kerr-type nonlinear material characterized by the 
third-order nonlinear susceptibility $\chi^{(3)}$
and the linear dielectric constant $\varepsilon_d^{(0)}$. 
For definiteness, we assume that 
$\varepsilon_d^{(0)} = \varepsilon_0 = 11.56$ and 
that the nonlinear defect rods are 
embedded into the photonic crystal along a selected 
direction $\vec{s}_{ij}$, so that they are located at the points 
$\vec{x}_m = \vec{x}_0 + m \, \vec{s}_{ij}$. 
As we show below, changing the radius $r_d$ of these defect 
rods and their location $\vec{x}_0$ in the crystal, one 
can create nonlinear waveguides with quite different properties.

\begin{figure}[t]
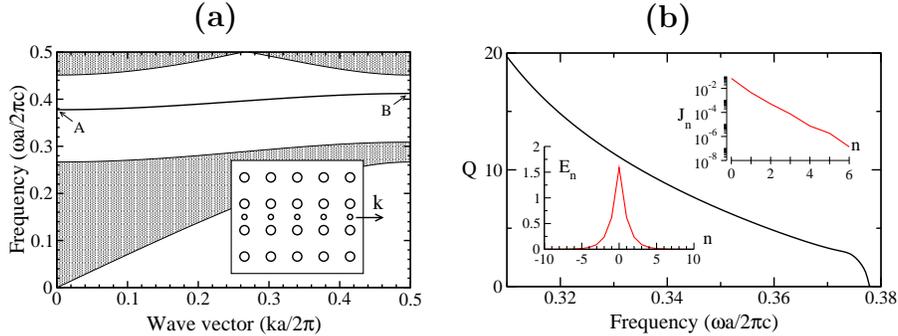

\begin{minipage}{55mm}
\centerline{{\large\bf (a)}} \vspace{1mm}
\includegraphics[width=55mm,angle=0,clip]{mk-fig5a.eps}
\end{minipage} \hspace*{3mm}
\begin{minipage}{55mm}
\centerline{{\large\bf (b)}} \vspace{1mm}
\includegraphics[width=58mm,angle=0,clip]{mk-fig5b.eps}
\end{minipage}
\vspace{3mm}
\caption{{\bf (a)} 
Dispersion relation for the photonic crystal waveguide 
shown in the inset ($\varepsilon_0=\varepsilon_d=11.56$, 
$r_0=0.18a$, $r_d=0.10a$). The grey areas are the projected 
band structure of the perfect 2D photonic crystal. 
The frequencies at the indicated points are: 
$\omega_A=0.378 \times 2\pi c/a$ and 
$\omega_B=0.412  \times 2\pi c/a$. 
{\bf (b)} Mode power $Q(\omega)$ of the nonlinear mode 
excited in the corresponding photonic crystal waveguide. 
The right inset gives the dependence $J_n(\omega)$ 
calculated at $\omega=0.37 \times 2\pi c/a$. 
The left inset presents the profile of the corresponding 
nonlinear localized mode.}
\label{fig:x2-0.10}
\end{figure}

As we have already discussed \cite{mingaleev,mingaleev2}, 
the Green function $G(\vec{x}, \vec{y} \,|\, \omega)$ 
and,  consequently,  the coupling coefficients 
$J_{m}(\omega)$ are usually highly long-ranged functions. 
This can be seen directly from Fig. \ref{fig:green-r0.18} 
that shows a typical spatial profile of the Green function. 
As a consequence, the coupling coefficients $J_n(\omega)$ calculated  
from Eq. (\ref{sys:Jn}) decrease slowly with the site number 
$n$. For $\vec{s}_{01}$ and $\vec{s}_{10}$ directions, the 
coupling coefficients can be approximated by an exponential 
function as follows 
\begin{displaymath}
|J_n(\omega)| \approx \left\{
\begin{array}{lcc}
J_0(\omega) \; , & \mbox{for} & n=0 \; , \\
J_{*}(\omega) \, e^{-\alpha(\omega) |n|} \; ,
& \mbox{for} & |n| \geq 1 \; ,
\end{array}
\right.
\end{displaymath}
where the characteristic decay rate $\alpha(\omega)$ can 
be as small as $0.85$, depending on the values of 
$\omega$, $\vec{x}_0$, and $r_d$, and 
it can be even smaller for other types of photonic 
crystals (for instance, for the photonic crystal used in 
Fig. \ref{fig:norm} we find 
$J_{m} \sim \, (-1)^m \exp(-0.66 m)$ for $m \geq 2$).
By this means, 
Eq. (\ref{sys:eq-E-disc}) is a nontrivial 
long-range generalization of a 2D discrete 
nonlinear Schr\"odinger (NLS) 
equation extensively studied during the last 
decade for different applications \cite{DNLS}. 
It allows us to draw an analogy between the problem
under consideration and a class of the NLS equations 
that describe  nonlinear excitations in quasi-one-dimensional molecular 
chains with long-range (e.g. dipole-dipole) interaction 
between the particles and local on-site nonlinearities 
\cite{Gaididei:1997:PRE,Johansson:1998:PRE}. 
For such systems, it was shown 
that the effect of nonlocal interparticle interaction 
brings some new features to the properties of nonlinear localized modes 
(in particular, bistability in their spectrum). 
Moreover, 
for our model the coupling coefficients $J_n(\omega)$ can be 
either unstaggered and monotonically decaying, i.e. 
$J_n(\omega)=|J_n(\omega)|$,  or 
staggered and oscillating from site to site, i.e. 
$J_n(\omega)=(-1)^{n}|J_n(\omega)|$.
We therefore 
expect that effective nonlocality in both linear and 
nonlinear terms of Eq. (\ref{sys:eq-E-disc}) may also 
bring similar new features into the properties of nonlinear 
localized modes excited in the photonic crystal
waveguides. 

\subsection{Staggered and unstaggered  localized modes}

As can be seen from the structure of the Green function 
presented in Fig. \ref{fig:green-r0.18}, the case of 
monotonically varying coefficients $J_n(\omega)$ can 
occur for the waveguide oriented in the $\vec{s}_{01}$ 
direction with $\vec{x}_0=\vec{a}_1/2$. In this case, the 
frequency of a linear guided mode, that can be excited in 
such a waveguide, takes a minimum value at $k=0$ [see 
Fig. \ref{fig:x2-0.10}(a)], and the corresponding 
nonlinear mode is expected to be unstaggered.

\begin{figure}[t]
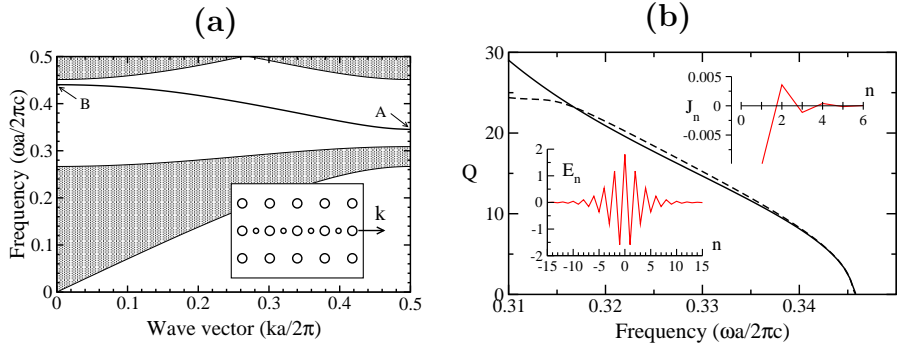

\begin{minipage}{55mm}
\centerline{{\large\bf (a)}} \vspace{1mm}
\includegraphics[width=55mm,angle=0,clip]{mk-fig6a.eps}
\end{minipage} \hspace*{3mm}
\begin{minipage}{55mm}
\centerline{{\large\bf (b)}} \vspace{1mm}
\includegraphics[width=58mm,angle=0,clip]{mk-fig6b.eps}
\end{minipage}
\vspace{3mm}
\caption{{\bf (a)} 
Dispersion relation for the photonic crystal waveguide 
shown in the inset ($\varepsilon_0=\varepsilon_d=11.56$, 
$r_0=0.18a$, $r_d=0.10a$). The grey areas are the projected 
band structure of the perfect 2D photonic crystal. 
The frequencies at the indicated points are: 
$\omega_A=0.346 \times 2\pi c/a$ and 
$\omega_B=0.440 \times 2\pi c/a$. 
{\bf (b)} Mode power $Q(\omega)$ of the nonlinear mode 
excited in the corresponding photonic crystal waveguide. 
Two cases are presented: the case of nonlinear rods in 
a linear photonic crystal ({\em solid line}) and the case 
of a completely nonlinear photonic crystal ({\em dashed 
line}). The right inset shows the behavior of the 
coupling coefficients $J_n(\omega)$ for 
$n \geq 1$ ($J_0=0.045$) at $\omega=0.33 \times 2\pi c/a$.
The left inset shows the profile of the  
nonlinear mode.}
\label{fig:x1-0.10}
\end{figure}

We have solved Eq. (\ref{sys:eq-E-disc}) numerically and found 
that nonlinearity can lead to the existence of {\em guided 
modes localized in both directions}, i.e. in 
the direction perpendicular to the waveguide, due to the 
guiding properties of a channel waveguide created by 
defect rods, and in the direction of the waveguide, 
due to the nonlinearity-induced self-trapping effect. 
Such nonlinear modes exist with the frequencies below the 
frequency of the linear guided mode of the waveguide, i.e. 
below the frequency $\omega_A$ in Fig. \ref{fig:x2-0.10}(a), 
and are indeed unstaggered, with the bell-shaped profile 
along the waveguide direction shown in the left inset of 
Fig. \ref{fig:x2-0.10}(b).

The 2D nonlinear modes localized in both dimensions can be 
characterized by the mode power which we define, by 
analogy with the NLS equation, as
\begin{equation}
Q = \sum_n |E_n|^2.
\label{sys:norm}
\end{equation}
This power is closely related to the energy of
the electric field in the 2D photonic crystal accumulated 
in the nonlinear mode. In Fig. \ref{fig:x2-0.10}(b) we 
plot the dependence of $Q$ on frequency, for the 
waveguide geometry shown in Fig. \ref{fig:x2-0.10}(a).

\begin{figure}[t]
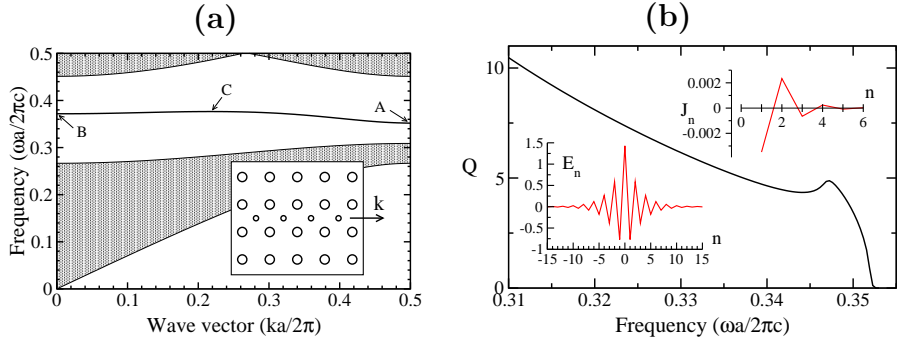

\begin{minipage}{55mm}
\centerline{{\large\bf (a)}} \vspace{1mm}
\includegraphics[width=55mm,angle=0,clip]{mk-fig7a.eps}
\end{minipage} \hspace*{3mm}
\begin{minipage}{55mm}
\centerline{{\large\bf (b)}} \vspace{1mm}
\includegraphics[width=58mm,angle=0,clip]{mk-fig7b.eps}
\end{minipage}
\vspace{3mm}
\caption{{\bf (a)} 
Dispersion relation for the photonic crystal waveguide 
shown in the inset ($\varepsilon_0=\varepsilon_d=11.56$, 
$r_0=0.18a$, $r_d=0.10a$). The grey areas are the projected 
band structure of the perfect 2D photonic crystal. 
The frequencies at the indicated points are: 
$\omega_A=0.352 \times 2\pi c/a$, $\omega_B=0.371 \times 
2\pi c/a$, and $\omega_C=0.376 \times 2\pi c/a$ (at $k=0.217 
\times 2\pi/a$).
{\bf (b)} Mode power $Q(\omega)$ of the nonlinear mode 
excited in the corresponding photonic crystal waveguide. 
The right inset shows the behavior of the 
coupling coefficients $J_n(\omega)$ for 
$n \geq 1$ ($J_0=0.068$) at  $\omega=0.345 \times 2\pi c/a$.
The left inset shows the profile of the corresponding 
nonlinear mode.}
\label{fig:x12-0.10}
\end{figure}

As can be seen from the Green function shown in 
Fig. \ref{fig:green-r0.18}, the case of staggered 
coupling coefficients $J_n(\omega)$ can be obtained for 
the waveguide oriented in the $\vec{s}_{10}$ direction 
with $\vec{x}_0=\vec{a}_1/2$.
In this case, the frequency dependence of the linear 
guided mode of the waveguide takes the minimum at 
$k=\pi/a$ [see Fig. \ref{fig:x1-0.10}(a)]. 
Accordingly, the nonlinear guided mode localized along the 
direction of the waveguide is expected to exist with the 
frequency below the lowest frequency $\omega_A$ of the 
linear guided mode, with a staggered profile. 
The longitudinal profile of such a 2D 
nonlinear localized mode is shown in the left inset in 
Fig. \ref{fig:x1-0.10}(b), together with the dependence of the 
mode power $Q$ on the frequency (solid curve), which in 
this case is again monotonic.

\begin{figure}
\begin{minipage}{63mm}
\centerline{{\large\bf (a)}}
\includegraphics[width=57mm,angle=0,clip]{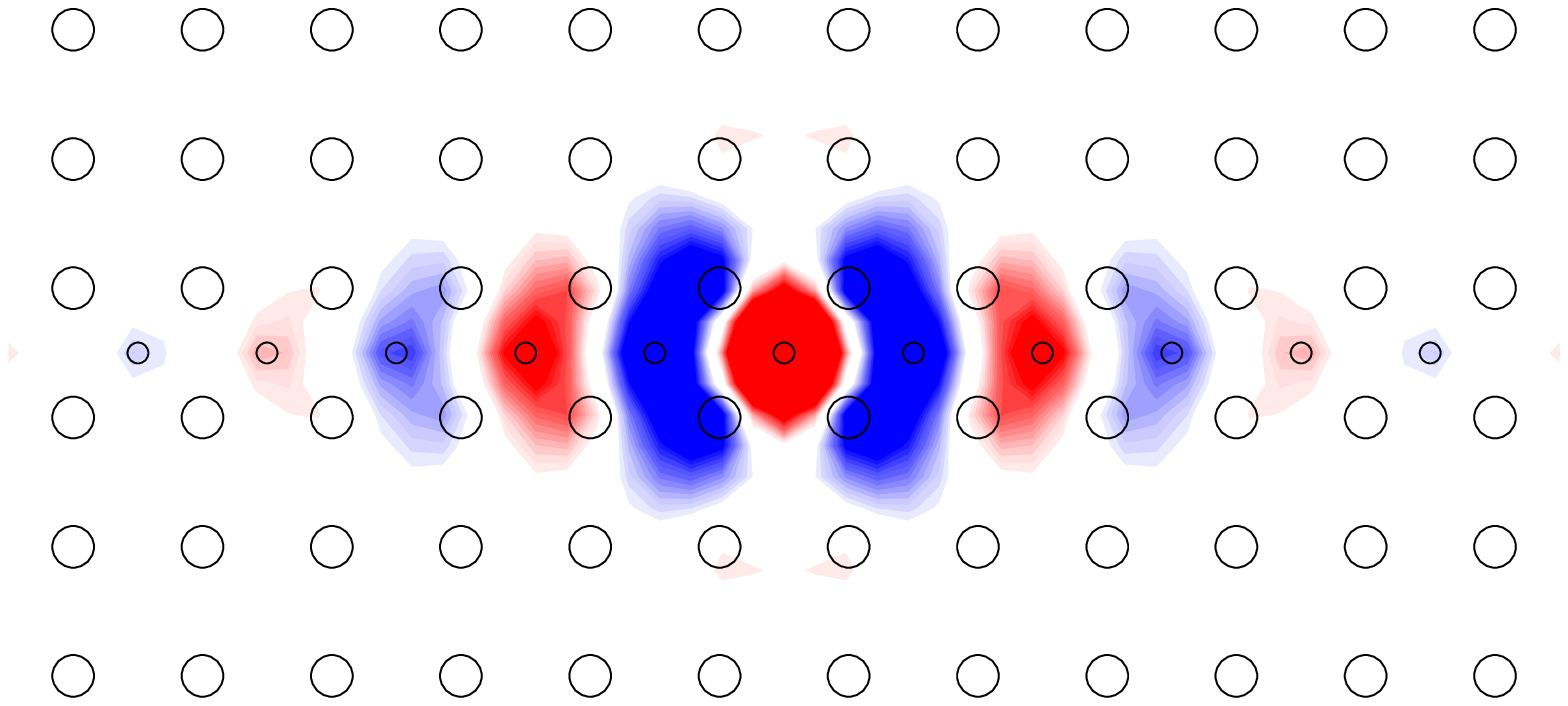}
\end{minipage} 
\hspace*{1mm}
\begin{minipage}{60mm}
\centerline{{\large\bf (b)}}
\includegraphics[width=53mm,angle=0,clip]{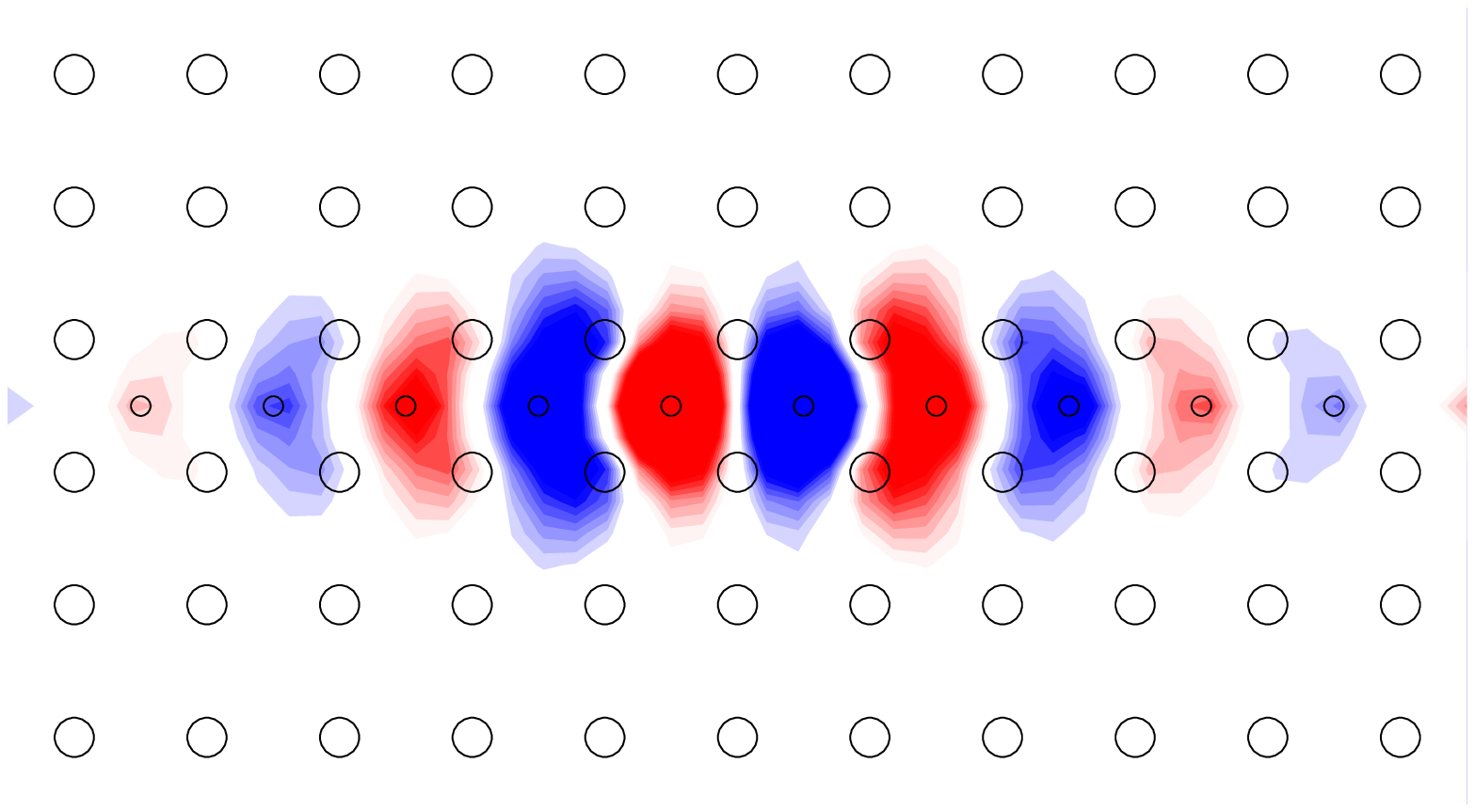}
\end{minipage}
\begin{minipage}{110mm}
\vspace{3mm}
\centerline{
\includegraphics[width=90mm,angle=0,clip]{mk-fig8c.eps}}
\end{minipage}
\vspace{3mm}
\caption{Examples of the {\bf (a)} symmetric and 
{\bf (b)} antisymmetric localized modes. 
The rod positions are indicated by circles and the amplitude 
of the electric field is indicated by color [red, for 
positive values, and blue, for negative values]; 
{\bf (c)} Power $Q$ vs. frequency dependencies 
calculated for two modes of different symmetry in the 
photonic crystal waveguide shown in Fig. 
\protect\ref{fig:x12-0.10}.}
\label{fig:antisymm}
\end{figure}

The results presented above are obtained for linear photonic 
crystals with nonlinear waveguides created by a row of defect 
rods. However, we have carried out the same analysis for the 
general case of {\em a nonlinear photonic crystal} that is 
created by rods of different size but made of the same 
nonlinear material. 
Importantly, we have found relatively small difference in all the 
results presented above provided nonlinearity is  
weak. In particular, for the photonic crystal waveguide 
shown in Fig. \ref{fig:x1-0.10}(a), the results for linear 
and nonlinear photonic crystals are very close. Indeed, for 
the mode power $Q$ the results corresponding to a nonlinear 
photonic crystal are shown in Fig. \ref{fig:x1-0.10}(b) by a 
dashed curve, and for $Q<20$ this curve almost coincides 
with the solid curve corresponding to the case of a 
nonlinear waveguide embedded into a 2D linear photonic 
crystal.

\subsection{Instability of nonlinear localized 
modes}

Let us now consider the waveguide created by a row of 
defect rods which are located at the points
$\vec{x}_0=(\vec{a}_1+\vec{a}_2)/2$, 
along a straight line in either the $\vec{s}_{10}$ or 
$\vec{s}_{01}$ directions. The results for this case are 
presented in Figs.
\ref{fig:x12-0.10}--\ref{fig:antisymm}. 
The coupling coefficients $J_n$ are described by a slowly 
decaying staggered function of the site number $n$, so that 
the effective interaction decays on the scale larger than 
in the two cases considered above. 

It is remarkable that, similar to the NLS models with 
long-range dispersive interactions 
\cite{Gaididei:1997:PRE,Johansson:1998:PRE}, 
we find a {\em non-monotonic} behavior of the mode power 
$Q(\omega)$ for this type of nonlinear photonic crystal 
waveguides: specifically, $Q(\omega)$ {\em increases} in 
the frequency interval $0.344 < (\omega a/ 2\pi c) < 0.347$ 
[shaded in Fig.~\ref{fig:antisymm}(c)]. One can expect 
that, similar to the results earlier obtained for the 
nonlocal NLS models 
\cite{Gaididei:1997:PRE,Johansson:1998:PRE}, 
the nonlinear localized modes in this interval 
are unstable and will eventually decay or 
transform into the modes of higher or lower frequency 
\cite{citeNLS}. What counts is that there is an interval 
of mode power in which {\em two stable nonlinear localized
modes of different widths do coexist}. Since the mode
power is closely related to the mode energy, one can
expect that the mode energy is also non-monotonic function
of $\omega$. Such a phenomenon is known as 
{\em bistability}, and in the problem under consideration 
it occurs as a direct manifestation of the nonlocality of 
the effective (linear and nonlinear) interaction between 
the defect rod sites. 

Being interested in the mobility of the nonlinear localized 
modes we have investigated, in addition to the symmetric modes 
shown in the left inset in Fig. \ref{fig:x12-0.10}(b) and 
in Fig. \ref{fig:antisymm}(a), 
also the {\em antisymmetric localized modes} shown in 
Fig. \ref{fig:antisymm}(b). 
Our calculations show that the power $Q(\omega)$ of the 
antisymmetric modes always (for all values of $\omega$ 
and all types of waveguides) exceeds that for symmetric 
ones [see, e.g., Fig. \ref{fig:antisymm}(c)]. 
Thus, antisymmetric modes are expected to be unstable and 
they should transform into a lower-energy symmetric 
modes. 

In fact, the difference between the power of antisymmetric and
symmetric modes determines the Peierls-Nabarro
barrier which should be overtaken for realizing 
the mobility of a nonlinear localized mode. One can see in Fig. 
\ref{fig:antisymm}(c) that the Peierls-Nabarro barrier is 
negligible for $0.347 < (\omega a/ 2\pi c) < 0.352$ and thus 
such localized modes should be mobile. However, the 
Peierls-Nabarro barrier becomes sufficiently large for 
highly localized modes with 
$\omega < 0.344 \times 2 \pi c/a$ and, as a consequence, 
such modes should be immobile. 
Hence, the bistability phenomenon in the photonic crystal
waveguides of the type depicted in Figs. 
\ref{fig:x12-0.10}--\ref{fig:antisymm}
opens up fresh opportunities \cite{Johansson:1998:PRE} 
for {\em switching} between immobile localized modes 
(used for the energy storage) and mobile localized modes 
(used for the energy transport).

The foregoing discussions on the mode mobility, based 
on the qualitative picture of the Peierls-Nabarro barrier, 
have been established for the discrete {\em one-dimensional} arrays. 
It is clear that the {\em two-dimensional} geometry of photonic 
crystals under consideration 
will bring new features into this picture. However, 
all these issues are still open and would require a further 
analysis.

\section{Self-Trapping of Light in a Reduced-Symmetry 
2D Nonlinear Photonic Crystal}

\begin{figure}[t]
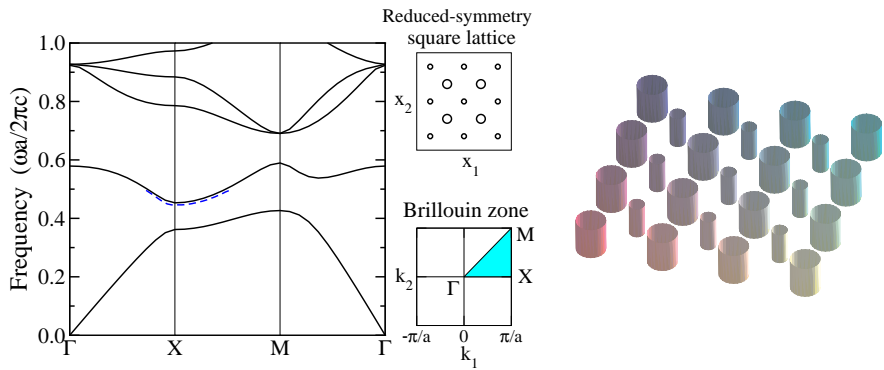

\begin{minipage}{70mm}
\includegraphics[width=70mm,angle=0,clip]{mk-fig9a.eps}
\end{minipage}
\begin{minipage}{35mm}
\includegraphics[width=35mm,angle=270,clip]{mk-fig9b.eps}
\end{minipage} 
\caption{Band-gap structure of the reduced-symmetry photonic 
crystal with $r_0 = 0.1a$, $r_d = 0.05a$, and 
$\varepsilon = 11.4$ for both types of rods. 
Full lines are calculated by the MIT Photonic-Bands 
program \protect\cite{mpb-prog} whereas dashed line is 
found from the effective discrete model.
The top center inset shows a cross-sectional 
view of the 2D photonic crystal depicted in the right 
inset. The bottom center inset shows the corresponding 
Brillouin zone.}
\label{fig:band}
\end{figure}

A low-intensity light cannot propagate through a photonic 
crystal if the light frequency falls into a band gap. 
However, it has been recently suggested \cite{john} that in 
the case of a 2D periodic medium with a Kerr-type nonlinear 
material, high-intensity light with frequency inside the 
gap can propagate in the form of {\em finite 
energy solitary waves} --  {\em 2D gap solitons}. 
These solitary waves were found to be {\em stable} 
\cite{john}, but the conclusion was based on the 
coupled-mode equations valid for a {\em weak modulation} of 
the dielectric constant $\varepsilon(\vec{x})$. 
However, in real photonic crystals the 
modulation of $\varepsilon(\vec{x})$ is {\em comparable to 
its average value}. 
Thus, the results of Ref. \cite{john} have a limited applicability to
the properties of localized modes in {\em realistic photonic crystals}. 

More specifically, the coupled-mode equations are valid if 
and only if the band gap $\Delta$ is vanishingly small, 
i. e. $\Delta \sim A^2$ where $A$ is an effective amplitude 
of the mode, that is a small
parameter in the multi-scale asymptotic expansions \cite{kiv}. 
If we apply this
model to describe nonlinear modes in a wider gap (see, e.g.,
discussions in Ref. \cite{kiv}), we obtain a 2D nonlinear
Schr{\"o}dinger (NLS) equation 
known to possess {\em no stable localized solutions}. 
Moreover, the 2D localized modes described by the coupled-mode equations 
are expected to possess
{\em an oscillatory instability} recently discovered 
for a broad class of coupled-mode Thirring-like
equations \cite{dima}. Thus, it is clear that, if nonlinear 
localized modes do exist in realistic PBG materials, their 
stability should be associated with {\em different physical 
mechanisms} not accounted for by simplified continuum
models. 

\begin{figure}
\centerline{
\includegraphics[width=80mm,angle=0,clip]{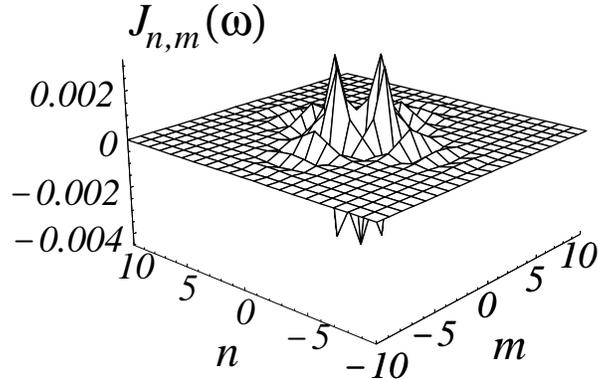}}
\vspace{3mm}
\caption{Coupling coefficients $J_{n,m}(\omega)$ for 
the photonic crystal depicted in Fig. 
\protect\ref{fig:band} (the contribution of the 
coefficient $J_{0,0}=0.039$ is not shown). 
The frequency $\omega=0.4456$ 
falls into the first band gap.}
\label{fig:Jnm}
\end{figure}

In this Section we follow Ref. \cite{mingaleev2} and 
study the properties of nonlinear
localized modes in a 2D photonic crystal composed of 
{\em two types of circular rods}: the rods of radius $r_0$ made
from a linear dielectric material and placed at the corners 
of a square lattice with the lattice spacing $a$, and the 
rods of radius $r_d$ made from a nonlinear dielectric 
material and placed at the center of each unit cell (see 
right inset in Fig. \ref{fig:band}). 
Recently, such {\em photonic 
crystals of reduced symmetry} have attracted considerable
interest because of their ability to possess 
{\em larger absolute band gaps} \cite{symmetry}. 
The band-gap structure of the 
reduced-symmetry photonic crystal is shown in 
Fig. \ref{fig:band}.
As is seen, it possesses two band gaps, first of
which extends from $\omega=0.426 \times 2 \pi c/a$ to 
$\omega=0.453 \times 2 \pi c/a$. 

\begin{figure}
\begin{minipage}{50mm}
\includegraphics[width=40mm,angle=0,clip]{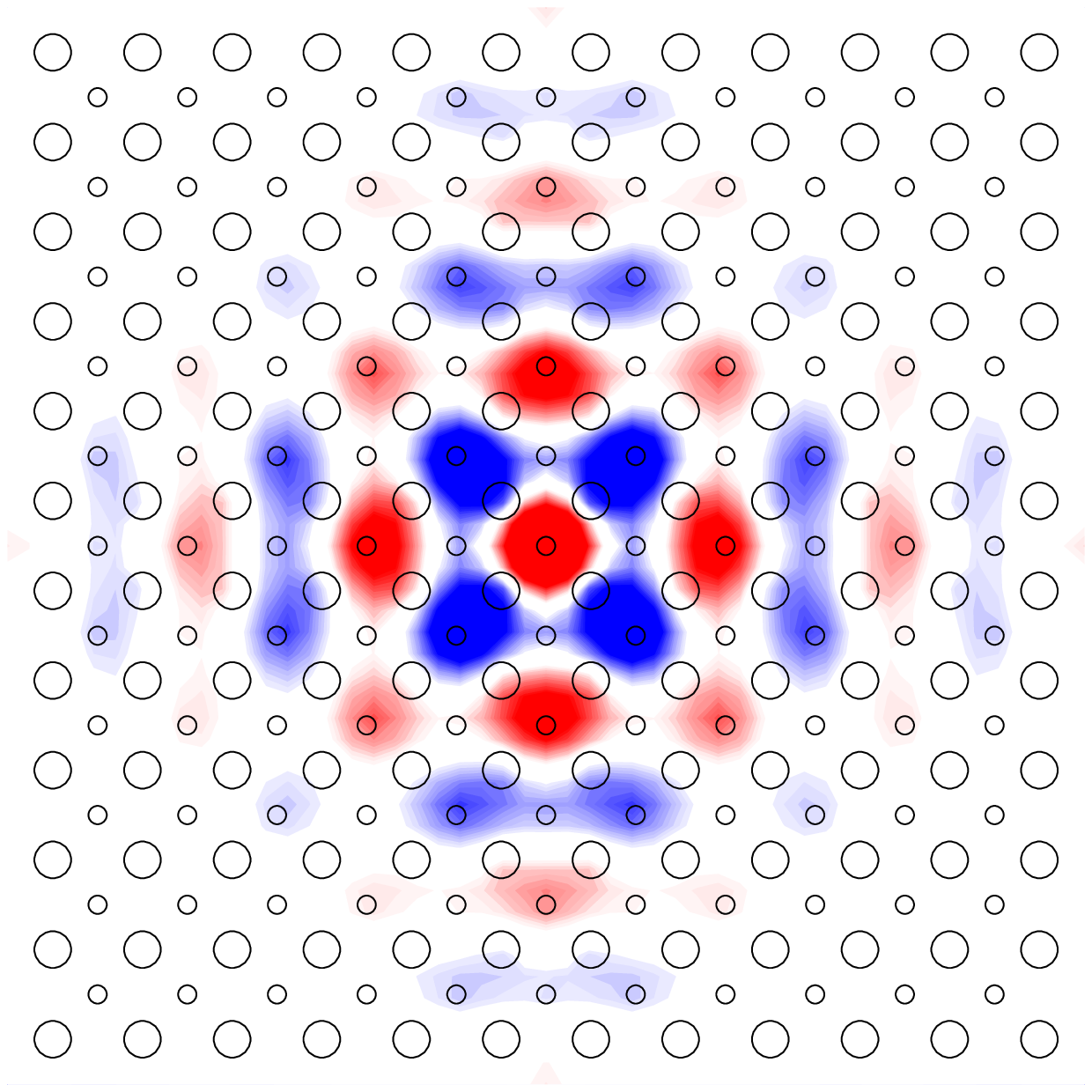}
\end{minipage} \hspace*{-5mm}
\begin{minipage}{50mm}
\includegraphics[width=70mm,angle=0,clip]{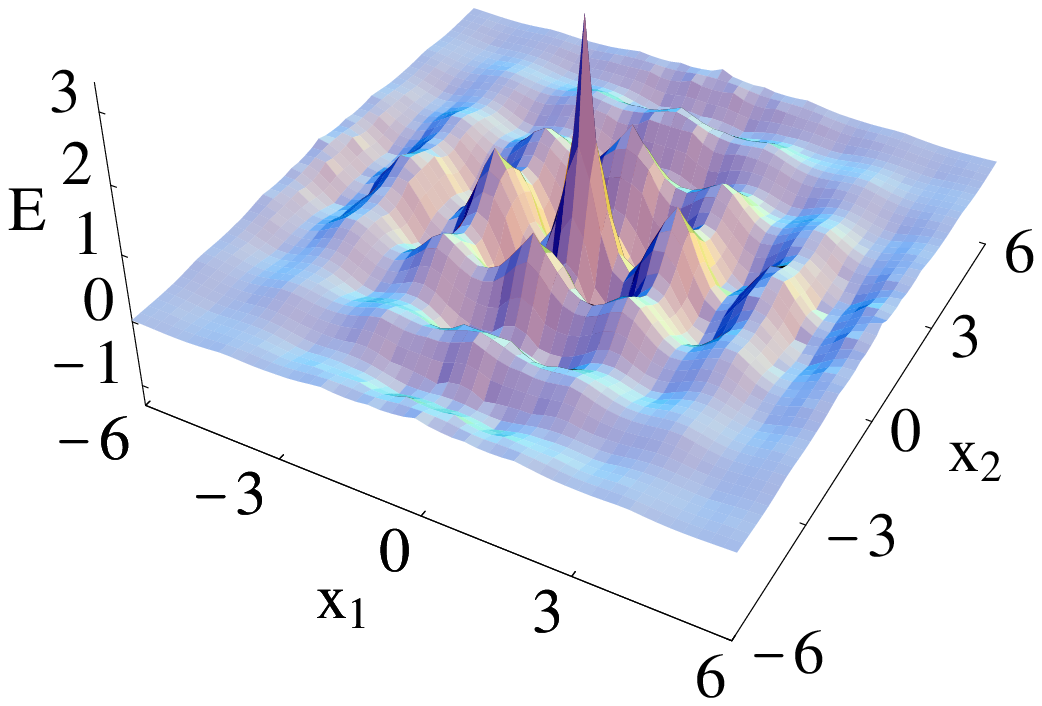}
\end{minipage}
\caption{Top (left) and 3D (right) views of a 
nonlinear localized mode in the first band gap of 
2D photonic crystal depicted in 
Fig. \protect\ref{fig:band}.}
\label{fig:mode}
\end{figure}

The reduced-symmetry ``diatomic'' photonic crystal shown in 
Fig. \ref{fig:band} can be considered as 
a square lattice of the ``nonlinear defect rods'' of small 
radius $r_d$ ($r_d < r_0$) embedded into the ordinary 
single-rod photonic crystal formed by a square lattice of 
rods of larger radius $r_0$ in air. 
The positions of the defect rods can then be described by the
vectors $\vec{x}_{n,m} = n \, \vec{a}_{1} + m \, 
\vec{a}_{2}$, where $\vec{a}_1$ and $\vec{a}_2$
are the primitive lattice vectors 
of the 2D photonic crystal. Here, in contrast to the 
photonic crystal waveguides discussed in the previous 
section, the nonlinear defect rods are characterized by
two integer indices, $n$ and $m$. However, it is
straightforward to extend Eq. (\ref{sys:eq-E-disc}) 
and write an approximate 2D discrete nonlinear equation 
\begin{eqnarray}
\label{sys:eq-E-disc:2D}
i \sigma \frac{\partial}{\partial t} E_{n,m} 
- E_{n,m} + \sum_{k,l} J_{n-k, \, m-l}(\omega) 
(\varepsilon_{d}^{(0)} + |E_{k,l}|^2) E_{k,l} 
 = 0\; ,
\end{eqnarray}
for the amplitudes of 
the electric field $E_{n,m}(t \,|\, \omega) \equiv 
E(\vec{x}_{n,m}, t \,|\, \omega)$ inside the 
defect rods. 
We have checked the accuracy of 
the approximation provided by Eq. (\ref{sys:eq-E-disc:2D}) 
solving it in the linear limit, in order to find the band-gap structure 
associated with linear stationary mode. 
Since the coupling coefficients $J_{n,m}(\omega)$ in the 
photonic crystal depicted in Fig. \ref{fig:band} 
are highly long-ranged functions 
(see Fig. \ref{fig:Jnm}), one should take into account the
interaction between at least 10 neighbors to reach accurate
results. 
As is seen from Fig. \ref{fig:band}, in this case 
the frequencies of the linear modes 
(depicted by a dashed line, with a minimum at 
$\omega=0.446 \times 2\pi c/a$) calculated from Eq. (\ref{sys:eq-E-disc:2D}) 
are in a good agreement with those calculated directly from  Eq.
(\ref{sys:eq-E-omega-t}). It lends a support to the validity 
of Eq. (\ref{sys:eq-E-disc:2D}) and allows us to use it for 
studying nonlinear properties. 

\begin{figure}[t]
\centerline{\hbox{
\includegraphics[width=90mm,angle=0,clip]{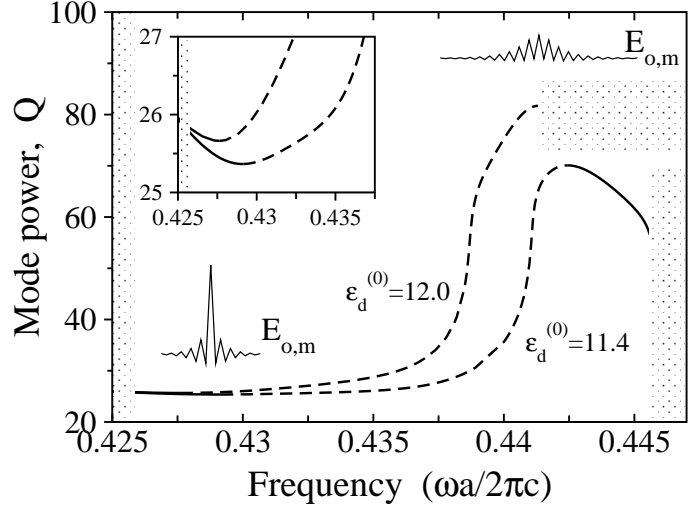}}}
\caption{Power $Q$  vs. frequency $\omega$ for 
the 2D nonlinear localized modes in the photonic crystal 
of Fig. \protect\ref{fig:band} with two different 
$\epsilon_d^{(0)}$. Solid lines -- stable modes, 
dashed lines -- unstable modes. Insets show typical profiles
of stable modes, and an enlarged part of the power
dependence. Grey areas show the lower and upper bands of 
delocalized modes surrounding the band gap.}
\label{fig:norm}
\end{figure}

Stationary nonlinear modes described by Eq. 
(\ref{sys:eq-E-disc:2D}) 
are found numerically by the Newton-Raphson 
iteration scheme. We reveal the existence of {\em a continuous family of 
such modes},  and a typical example [smoothed by continuous optimization for 
Eq. (\ref{sys:eq-E-omega-t2})] of nonlinear localized mode 
is shown in Fig. \ref{fig:mode}. 
In Fig. \ref{fig:norm}, we plot the dependence 
of the mode power 
\begin{equation}
Q(\omega) = \sum_{n,m} |E_{n,m}|^2 \; ,
\label{sys:norm:2D}
\end{equation}
on the frequency $\omega$ for 
the photonic crystal shown in Fig. \ref{fig:band}. 
As we have already discussed, 
this dependence represents a very important characteristic of 
nonlinear localized modes which allows to determine 
their stability by means of the Vakhitov-Kolokolov 
stability criterion: $dQ/d\omega>0$ for unstable modes 
(this criterion has been extended \cite{Laedke} 
to 2D NLS models). 

As is well known \cite{Laedke,Stabil}, 
in the 2D discrete cubic NLS equation, only high-amplitude 
localized modes are stable, whereas no stable modes exist in
the continuum limit. For our model, the high-amplitude modes
are also stable (see inset in Fig. \ref{fig:norm}), 
but they are not accessible under realistic conditions: 
To excite such modes one  should increase the refractive index
at the mode center in more than 2 times. 
Thus, for realistic conditions and relatively small values of 
$\chi^{(3)}$, only low-amplitude localized modes become a 
subject of much interest since they can be excited in 
experiment. However, such modes in unbounded 2D NLS models 
are always unstable and either collapse or spread out 
\cite{DNLS}.  
In fact, they can be stabilized by some external forces 
(e.g., due to interactions with boundaries or 
disorder \cite{disorder}), but in this case the excitations 
are pinned and cannot be used for energy or signal transfer.

Here we reveal that, in a sharp contrast to the 
2D discrete NLS models discussed earlier in various 
applications, the low-amplitude localized modes 
of Eq. (\ref{sys:eq-E-disc:2D}) can be stabilized 
due to {\em nonlinear long-range dispersion} inherent to 
the photonic crystals.  It should be emphasized that such 
stabilization does not occur in the models with only {\em linear long-range} 
dispersion \cite{DNLS}. 
In order to gain a better insight into the stabilization 
mechanism, we have carried out the studies of Eq. 
(\ref{sys:eq-E-disc}) for the exponentially decaying 
coupling coefficients $J_{n,m}$. Our results show that 
the most important factor which determines stability 
of the low-amplitude localized modes is a ratio of the 
coefficients 
at the local nonlinearity ($\sim J_{0,0}$) and the 
nonlinear dispersion ($\sim J_{0,1}$). If the coupling 
coefficients $J_{n,m}$ decrease with the distances $n$ and $m$ 
rapidly, the low-amplitude modes of 
Eq. (\ref{sys:eq-E-disc:2D}) 
with $\epsilon_d^{(0)}=11.4$ are essentially stable 
for $J_{0,0}/J_{0,1} \leq 13$. However, this estimation 
is usually lowered because the stabilization is favored 
by the presence of long-range interactions. 

It should be mentioned that the stabilization of 
low-amplitude 2D localized modes is not inherent to all types 
of nonlinear photonic crystals. On the contrary, the photonic 
crystals must be {\em carefully designed} to support 
{\em stable low-amplitude nonlinear modes}. For example, in the
photonic crystal considered above such modes are stable 
at least for $11 < \epsilon_d^{(0)} < 12$, however they become 
unstable for $\epsilon_d^{(0)} \geq 12$ 
(see Fig. \ref{fig:norm}). 
The stability of these modes can also be controlled by 
varying $r_d$, $r_0$, or $\epsilon_0$.

\section{Concluding Remarks}

Exploration of nonlinear properties of PBG materials
may open new important application of photonic crystals 
for all-optical signal processing  and switching, 
 allowing an effective way to create tunable band-gap 
structures operating entirely with light. Nonlinear 
photonic crystals,  
and nonlinear waveguides created in the photonic structures 
with 
a periodically modulated dielectric constant, create an ideal 
environment for the generation and observation of nonlinear 
localized modes.

As follows from our results, nonlinear localized modes can 
be excited in photonic crystal 
waveguides of different geometry. For several geometries 
of 2D 
waveguides, we have demonstrated that such modes
are described by a new type of nonlinear lattice models 
that include 
long-range interaction and effectively nonlocal nonlinear 
response. 
It is expected that the general features of nonlinear guided modes 
described here will be preserved in other types of photonic crystal 
waveguides.  Additionally, similar types of nonlinear localized modes are 
expected in photonic crystal fibers \cite{russell} consisting of a 
periodic air-hole lattice that runs along the length of the fiber, 
provided the fiber core is made of a highly nonlinear material 
(see, e.g., Ref. \cite{egg}).

Experimental observation of nonlinear photonic localized modes would 
require not only the use of photonic materials with a relatively large 
nonlinear refractive index (such as AlGaAs waveguide PBG 
structures \cite{algas} or polymer PBG crystals \cite{jap}, but also 
a control of the group-velocity dispersion and band-gap parameters. 
The latter can be achieved by employing the surface coupling 
technique \cite{ast} that is able to provide coupling to specific 
points of the dispersion curve, opening up a very straightforward way 
to access nonlinear effects.

\section*{Acknowledgments}

The authors are indebted to O. Bang, K. Busch, 
P.L. Christiansen, Yu.B. Gaididei, S. John, A. McGurn, 
C. Soukoulis, and A.A. Sukhorukov for encouraging 
discussions, and R.A. Sammut for collaboration at 
the initial stage of this project. 
The work has been partially supported by the Large Grant 
Scheme of the Australian Research Council and the 
Performance and Planning Foundation grant of the 
Institute of Advanced Studies. 


\end{document}